# The Complexity of Rerouting Shortest Paths

Paul Bonsma

Computer Science Department, Humboldt University Berlin,
Unter den Linden 6, 10099 Berlin. `bonsma@informatik.hu-berlin.de`

**Abstract.** The Shortest Path Reconfiguration problem has as input a graph $G$ (with unit edge lengths) with vertices $s$ and $t$, and two shortest $st$-paths $P$ and $Q$. The question is whether there exists a sequence of shortest $st$-paths that starts with $P$ and ends with $Q$, such that subsequent paths differ in only one vertex. This is called a rerouting sequence. This problem is shown to be PSPACE-complete. For claw-free graphs and chordal graphs, it is shown that the problem can be solved in polynomial time, and that shortest rerouting sequences have linear length. For these classes, it is also shown that deciding whether a rerouting sequence exists between *all* pairs of shortest $st$-paths can be done in polynomial time. Finally, a polynomial time algorithm for counting the number of isolated paths is given.

## 1 Introduction

In this paper, we study the *Shortest Path Reconfiguration (SPR) Problem*, introduced by Kamiński et al [16,15]. The input consists of a graph $G$, with vertices $s$ and $t$, and two shortest $st$-paths $P$ and $Q$. The question is whether $P$ can be modified to $Q$ by changing one vertex at a time, and maintaining a shortest $st$-path throughout. Edges have unit lengths, so all shortest $st$-paths have the same number of vertices. We define the following *solution graph* $SP(G, s, t)$: its vertex set is the set of all shortest $st$-paths in $G$. Two paths $P$ and $Q$ are adjacent if they differ in one vertex. SPR can now be reformulated as: does there exist a walk from $P$ to $Q$ in $SP(G, s, t)$? Such a walk is also called a *rerouting sequence*.

Shortest paths form a central concept in graph theory, optimization, algorithms and networking. Questions related to rerouting (shortest) paths are often studied in networking applications. Hence this is a very natural question, which is likely to have relevant applications. Nevertheless, the main motivation for this study is of a more theoretical nature. Similar reconfiguration problems can be defined based on many different combinatorial problems: Consider all solutions to a problem (or all solutions of at least/at most given weight, in the case of optimization problems), and define a (symmetric) adjacency relation on them. Such problems have been studied often in recent literature. Examples include reconfiguration problems based on satisfiability problems [10], independent sets [11,13,17], vertex colorings [1,3,4,5,6], matchings [13], list edge-colorings [14], matroid bases [13], subsets of a (multi)set of numbers [9]. Of course, to obtain a reconfiguration problem, one needs to define an adjacency relation between solutions. Usually, the most natural adjacency relation is considered, e.g. two independent sets $I$ and $J$ are considered adjacent in [13] if $J$ can be obtained from $I$ by removing one vertex and adding another; boolean assignments are considered adjacent in [10] if exactly one variable differs, etc. We remark that in the context of local search, similar problems have been studied earlier, with the important distinction that the neighborhood is not symmetric, and the objective is to reach a local optimum, instead of a given target solution, see e.g. [18].

An initial motivation of these questions was to explore the solution space of NP-hard problems, to study e.g. the performance of heuristics [10] and random sampling methods [4]. This has revealed interesting, often recurring patterns in the complexity behavior of these problems. This is perhaps best exemplified by the known results on the reconfiguration of vertex colorings using $k$ colors: in the problem $k$-Color Path, two $k$-colorings of a graph are given, and the question is whether one can be modified to the other by changing one vertex color at a time, and maintaining a $k$-coloring throughout. This problem is polynomial time solvable for $k \leq 3$ [6], and PSPACE-complete for $k \geq 4$ [3]. Note that the corresponding decision problem of deciding whether a graph admits a $k$-coloring is polynomial time solvable for $k \leq 2$, and NP-complete for $k \geq 3$. This gives an example of the following common pattern: for instance classes for which deciding whether a solution exists is in P, the reconfiguration problem is often in P as well. See [10,12,13] for more extensive examples. This motivated Ito et al [12] to ask for examples of reconfiguration problems that break this pattern. Secondly, it has been observed that there is a strong correlation between the complexity of reconfiguration problems and the diameter of the components of the solution graph: for all known 'natural' reconfiguration problems in P, the diameter is polynomially bounded (see e.g. [1,6,10,13,17]), and for all PSPACE-complete reconfiguration problems, the diameter may be superpolynomial or exponential (see e.g. [3,10]). The latter is unsurprising, since polynomial diameter would imply NP=PSPACE (assuming that the property of being a solution and adjacency of solutions can be tested in polynomial time, which holds for all aforementioned problems). One can easily construct artificial instance classes of reconfiguration problems such that the problem is in P, but has exponential diameter [3], but to our knowledge no natural examples are known. (That is, not constructed specifically to prove something about the reconfiguration problem at hand.)

With the goal of breaking one of these patterns, Kamiński et al [15,16] introduced the SPR problem. Finding a shortest path can be done in polynomial time. Nevertheless, in [15,16] examples were constructed where the solution graph has exponential diameter. This shows that regardless of whether SPR is in P or PSPACE-complete, one of the patterns is broken. The main open question from [16] was therefore that of determining the complexity of SPR.

In this paper, we answer that question by showing that SPR is PSPACE-complete. Therefore, this also answers the question posed in [12], by giving a rare example of a PSPACE-complete reconfiguration problem based on a decision problem in P. We remark that it is not the first example: in [3] it is shown that 4-Color Path is also PSPACE-hard for bipartite graphs. Since every bipartite graph is 2-colorable, the corresponding decision problem is trivial. Our PSPACE-completeness result is presented in Section 3. We remark that our PSPACE-completeness result, after it appeared in a preprint [2], has already proved its usefulness for showing PSPACE-completeness of other problems: in [17], the result has been applied to show that Independent Set Reconfiguration remains PSPACE-hard even when restricted to perfect graphs.

We give the following positive results on SPR. We show that when $G$ is chordal or claw-free, SPR can be decided in polynomial time. A graph is *chordal* if it contains no induced cycles of length more than 3. This is a well-studied class of perfect graphs, which includes for instance $k$-trees and interval graphs [8]. A graph is *claw-free* if it contains no induced $K_{1,3}$ subgraph. This is again a well-studied graph class, see e.g. [7]. We also show that for these graph classes, the diameter of components of $\mathrm{SP}(G, s, t)$ is always linearly bounded. For chordal graphs, we can actually construct a *shortest* rerouting sequence in polynomial time. In contrast, in [15], it was shown that for general graphs, finding a shortest rerouting sequence is NP-hard, even for graph classes where there always exists one of polynomial length.

In the context of reconfiguration problems, other types of questions are commonly studied as well. Above, we considered the *reachability question*: can one given solution be reached from another given solution? The related *connectivity question* has also been well-studied [10,4,5,9]: is the solution graph connected? For chordal graphs $G$, we show that $\mathrm{SP}(G, s, t)$ is always connected. If $G$ is claw-free, we show that it can be decided in polynomial time whether $\mathrm{SP}(G, s, t)$ is connected. Our results on chordal graphs are presented in Section 4, and the results on claw-free graphs in Section 5.

Another type of question that has been studied in this context is related to the existence of isolated states [9]. In the case of SPR, an *isolated st-path* is a shortest $st$-path in $G$ that has no neighbors in $\mathrm{SP}(G, s, t)$. The reader may observe that deciding whether a given path is an isolated $st$-path is a trivial problem, that can be decided in linear time. Similarly, deciding whether all shortest $st$-paths are isolated can trivially be done in polynomial time as well. The problem of deciding whether there exists an isolated $st$-path is less trivial. In Section 6 we give an algorithm for this problem. In fact, we give a polynomial time algorithm for the more general problem of *counting* the number of isolated paths. Statements for which (more detailed) proofs can be found in the appendix are marked with a star.

## 2 Preliminaries

For graph theoretical notions not defined here, we refer to [8]. We will consider undirected and simple graphs throughout. A *walk* of length $k$ from $v_0$ to $v_k$ in a graph $G$ is a vertex sequence $v_0, \ldots, v_k$, such that for all $i \in \{0, \ldots, k-1\}$, $v_i v_{i+1} \in E(G)$. It is a *path* if all vertices are distinct. It is a *cycle* if $k \geq 3$, $v_0 = v_k$, and $v_0, \ldots, v_{k-1}$ is a path. With a path or cycle $W = v_0, \ldots, v_k$ we associate a subgraph of $G$ as well, with vertex set $V(W) = \{v_0, \ldots, v_k\}$ and edge set $E(W) = \{v_i v_{i+1} \mid i \in \{0, \ldots, k-1\}\}$. A path from $s$ to $t$ is also called an *st-path*. The *distance* from $s$ to $t$ is the length of a *shortest st-path*. The *diameter* of a graph is the maximum distance from $s$ to $t$ over all vertex pairs $s, t$.

A hypergraph $H = (V, E)$ consists of a vertex set $V$, and a set $E$ of hyperedges, which are subsets of $V$. A walk in $H$ of length $k$ is a sequence of vertices $v_0, \ldots, v_k$ such that for every $i$, there exists a hyperedge $e \in E$ with $\{v_i, v_{i+1}\} \subseteq e$. Using this notion of walks, connectivity and components of hypergraphs are defined the same as for graphs.

Throughout this paper, we will consider a graph $G$ with vertices $s, t \in V(G)$. We will only be interested in shortest $st$-paths in $G$, and use $d$ to denote their length. For $i \in \{0, \ldots, d\}$, we define $L_i \subseteq V(G)$ to be the set of vertices that lie on a shortest $st$-path, at distance $i$ from $s$. So $L_0 = \{s\}$, and $L_d = \{t\}$ (even if there may be more vertices at distance $d$ of $s$). A set $L_i$ is also called a *layer*. With respect to a given layer $L_i$, the *previous layer* is $L_{i-1}$, and the *next layer* is $L_{i+1}$. Clearly, if there is an edge $xy \in E(G)$ with $x \in L_i$ and $y \in L_j$, then $|j - i| \leq 1$. Note that a shortest $st$-path $P$ contains exactly one vertex from every layer. For $i \in \{0, \ldots, d\}$, this vertex will be called the $L_i$-*vertex* of $P$.

The graph $G$ will be undirected, so we use the notation $N(v)$ to denote the set of neighbors of a vertex $v \in V(G)$. However, if $v \in L_i$, then we will use $N^-(v)$ to denote $N(v) \cap L_{i-1}$, and call these neighbors the *in-neighbors of* $v$. Similarly, $N^+(v)$ denotes $N(v) \cap L_{i+1}$, and these are called the *out-neighbors of* $v$.

Recall that a rerouting sequence from $P$ to $Q$ is a sequence $Q_0, \ldots, Q_k$ of shortest $st$-paths with $Q_0 = P$, $Q_k = Q$, such that for every $j \in \{0, \ldots, k-1\}$, $Q_j$ and $Q_{j+1}$ differ in exactly one vertex. Let $L_i$ be the layer in which they differ, and $u = L_i \cap V(Q_j)$ and $v = L_i \cap V(Q_{j+1})$. Then we also say that $Q_{j+1}$ is obtained from $Q_j$ with a rerouting step $u \to v$ in layer $L_i$.

## 3  PSPACE-completeness

In this section we prove that the SPR problem is PSPACE-complete. A *k-color assignment* $\alpha$ for a graph $G$ is a function $\alpha : V(G) \to \{1, \ldots, k\}$. A *k-coloring* $\alpha$ for a graph $G$ is a color assignment such that for all $uv \in E(G)$, $\alpha(u) \neq \alpha(v)$. For a given graph $G$, the *k-color graph* $\mathcal{C}_k(G)$ has vertex set consisting of all $k$-colorings of $G$, where two colorings are adjacent if they differ only in one vertex. A walk in $\mathcal{C}_k(G)$ from $\alpha$ to $\beta$ will also be called a *recoloring sequence* from $\alpha$ to $\beta$. The problem *k-Color Path* has as input a graph $G$, with two $k$-colorings $\alpha$ and $\beta$. The question is whether there exists a walk from $\alpha$ to $\beta$ in $\mathcal{C}_k(G)$. $k$-Color Path has been shown to be PSPACE-complete in [3].

Let $G$, $\alpha$, $\beta$ be an instance of 4-Color Path, with $V(G) = \{v_1, \ldots, v_n\}$. We will now describe how to construct an equivalent SPR instance $G'$ with two shortest $st$-paths $P_\alpha$ and $P_\beta$. Every shortest $st$-path in $G'$ will correspond to a 4-color assignment for $G$ (though not necessarily a 4-coloring!). To indicate this correspondence, some vertices of $G'$ will be colored with the four colors $\{1, 2, 3, 4\}$. The other vertices will be colored with a fifth color, namely *black*. Note that this 5-color assignment for $G'$ will not be a coloring of $G'$. $G'$ will consist of one *main strand*, which contains the paths $P_\alpha$ and $P_\beta$, and $6n$ *recoloring strands*: one for every combination of a vertex $v_i \in V(G)$ and two colors $\{c_1, c_2\} \subset \{1, 2, 3, 4\}$.

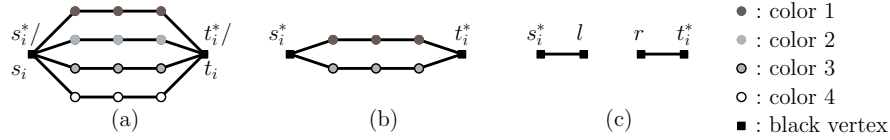

**Fig. 1.** Different variants of the gadgets $H_i$ and $H_i^*$ used in the construction.

The construction of $G'$ starts by introducing the vertices $s$ and $t$. The *main strand* is constructed as follows. For each $v_i \in V(G)$, introduce a vertex gadget $H_i$ as shown in Figure 1 (a). The leftmost vertex of $H_i$ is labeled $s_i$, and the rightmost vertex $t_i$. These vertices are colored black. $H_i$ consists of four disjoint $s_i t_i$-paths of length 4, one for each color. All internal vertices of the paths are colored in the color assigned to the path. The four vertices of $H_i$ that are neither adjacent to $s_i$ nor to $t_i$ are called *middle vertices* of $H_i$. These gadgets $H_i$ are connected as follows: add edges $ss_1$ and $t_n t$, and for every $i \in \{1, \ldots, n-1\}$, add an edge $t_i s_{i+1}$. At this point the graph is connected, and every vertex lies on a shortest $st$-path. Observe that the distance from $s$ to $s_i$ ($t_i$) is $5i - 4$ (resp. $5i$), and the distance from $s$ to $t$ is $5n + 1$. Hence this defines for every vertex the layer $L_i$ that it is part of, for $i \in \{0, \ldots, 5n + 1\}$.

We now define the *recoloring strands*. For each $v_i \in V(G)$ and each color pair $\{c_1, c_2\} \subset \{1, 2, 3, 4\}$, we introduce a recoloring strand called the $v_i, \{c_1, c_2\}$-*strand*, defined as follows. Let $\{c_3, c_4\} = \{1, 2, 3, 4\} \backslash \{c_1, c_2\}$. First we introduce gadgets $H_j^*$ for every $j \in \{1, \ldots, n\}$. (Whenever we mention gadgets $H_j^*$ or vertices $s_j^*, t_j^*, l$ and $r$ below, this refers to the gadgets and vertices for the given $v_i, \{c_1, c_2\}$-strand.)

- If $j \neq i$ and $v_i v_j \notin E(G)$, then define $H_j^*$ to be isomorphic to $H_j$ (see Figure 1 (a)), with the same 5-color assignment. The leftmost and rightmost (black) vertices are now labeled $s_j^*$ and $t_j^*$ respectively.
- If $j \neq i$ and $v_i v_j \in E(G)$, then define $H_j^*$ to be as shown in Figure 1 (b). The leftmost and rightmost (black) vertices are labeled $s_j^*$ and $t_j^*$ again. Now there are only two disjoint paths from $s_j^*$ to $t_j^*$, which are colored with the colors $c_3$ and $c_4$.
- $H_i^*$ is the gadget shown in Figure 1 (c). Here $s_i^*$ has one neighbor labeled $l$, and $t_i^*$ has one neighbor labeled $r$.

Complete the strand by adding edges $ss_1^*$, $t_n^* t$ and $t_j^* s_{j+1}^*$ for every $j \in \{1, \ldots, n-1\}$. Note that if we add edges from $l$ and $r$ to a main strand vertex in layer $L_{5i-2}$, which we will do below, then all vertices of the new strand lie on $st$-paths of length $5n + 1$ as well, and no shorter $st$-paths have been created. This defines for every vertex in the new strand which distance layer it is part of. We will refer to these layers in the next step, where we show how to connect the vertices of this recoloring strand to the main strand; see Figure 2. For all $j < i$, add the following edges: Add edges between $s_j^*$ and every main-strand vertex in the next layer that has a color that is also used in $H_j^*$. Next, for every non-black vertex $v$ of $H_j^*$, add an edge between $v$ and the main-strand vertex in the next layer that has the same color as $v$, or is black. Finally, add an edge $t_j^* s_{j+1}$. For all $j > i$, edges between $H_j^*$ and the main strand are added similarly, except that vertices of $H_j^*$ are connected to vertices in the *previous* layer. (See $H_4^*$ in Figure 2 for an example.) For $H_i^*$ we add edges as follows: Connect $s_i^*$ ($t_i^*$) to the main strand vertices in the next (resp. previous) layer with colors $c_1$ and $c_2$. Finally, connect both remaining vertices $l$ and $r$ of $H_i^*$ to both middle vertices of $H_i$ that have colors $c_1$ and $c_2$. Introducing such a $v_i, \{c_1, c_2\}$-*strand* for every $v_i \in V(G)$ and $\{c_1, c_2\} \subset \{1, 2, 3, 4\}$ completes the construction of $G'$.

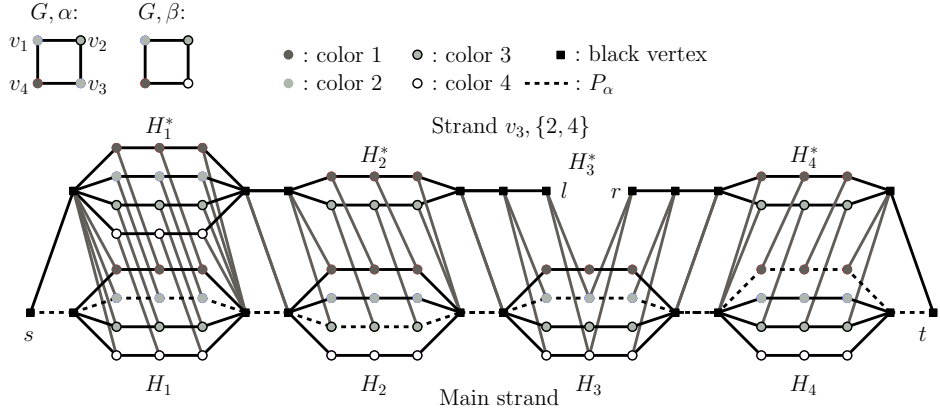

**Fig. 2.** A $k$-Color Path instance $G$, $\alpha$, $\beta$, and two strands of the resulting graph $G'$.

We now show how to construct a path $P_\gamma$ for any given 4-coloring $\gamma$ of $G$; see Figure 2. The path $P_\gamma$ contains only main strand vertices. Since it should be a shortest $st$-path, it contains exactly one vertex of every layer. Every layer contains vertices of a unique gadget $H_i$ of the main strand. In the case that the layer contains a single black vertex from $H_i$, this is the vertex that is included in $P_\gamma$. In the case that the layer contains vertices of colors $1, \ldots, 4$ of $H_i$, use the vertex of color $\gamma(v_i)$ for $P_\gamma$. This way, we define the paths $P_\alpha$ and $P_\beta$, using the given colorings $\alpha$ and $\beta$, respectively.

The purpose of the recoloring strands is as follows: Consider two adjacent colorings $\alpha$ and $\beta$. Suppose they differ only in vertex $v_i$, where their respective colors are $c_1 = \alpha(v_i)$ and $c_2 = \beta(v_i)$. Then all neighbors of $v_i$ are colored with colors in $\{1, 2, 3, 4\} \backslash \{c_1, c_2\}$. Therefore, $P_\alpha$ can be rerouted to a shortest $st$-path $P'$ that lies entirely in the $v_i, \{c_1, c_2\}$-strand, except for the vertex in layer $L_{5i-2}$. This rerouting is done by making rerouting steps in layers $L_1, \ldots, L_{5i-1}$ in increasing order, and subsequently in layers $L_{5n}, \ldots, L_{5i-3}$ in decreasing order. Note that the path $P'$ must use vertices that have the same color as the $P_\alpha$-vertex of the same layer. Then, with a single rerouting step, the color of the vertex in layer $L_{5i-2}$ can be changed from $c_1$ to $c_2$. Rerouting the path back to the main strand, in reversed layer order, then gives $P_\beta$. This can be done for every pair of adjacent colorings, so a recoloring sequence from $\alpha$ to $\beta$ gives a rerouting sequence from $P_\alpha$ to $P_\beta$. This yields:

**Lemma 1 (*)** *If there is a recoloring sequence for $G$ from $\alpha$ to $\beta$, then there is a rerouting sequence from $P_\alpha$ to $P_\beta$ for $G'$.*

With any shortest $st$-path $P'$ in $G'$, we associate a color assignment where $v_i \in V(G)$ receives the same color as the (middle) vertex in $L_{5i-2} \cap V(P')$. The converse of Lemma 1 can then be proved as follows: a rerouting step can only change the $L_{5i-2}$-vertex of a shortest $st$-path $P'$ from a vertex of color $c_1$ to a vertex of color $c_2$ if the vertices of $P'$ in the previous and next layer are part of the $v_i, \{c_1, c_2\}$-strand. In that case, all vertices of $P'$ except the $L_{5i-2}$-vertex must be part of this strand, which shows that $P'$ corresponds to a 4-color assignment in which the neighbors of $v_i$ are not colored with $c_1$ or $c_2$. Hence if $P'$ corresponds to a 4-coloring, then the shortest $st$-path resulting from the rerouting step corresponds again to a 4-coloring. Therefore, any rerouting sequence from $P_\alpha$ to $P_\beta$ can be mapped to a recoloring sequence from $\alpha$ to $\beta$.

**Lemma 2 (*)** *If there is a rerouting sequence from $P_\alpha$ to $P_\beta$ for $G'$, then there is a recoloring sequence in $G$ from $\alpha$ to $\beta$.*

**Theorem 3** *SPR is PSPACE-complete.*

*Proof:* 4-Color Path is PSPACE-complete [3]. Our transformation from $G$ to $G'$ is polynomial. By Lemma 1 and 2, $G$, $\alpha$, $\beta$ is a YES-instance for 4-Color Path if and only if $G'$, $P_\alpha$, $P_\beta$ is a YES-instance for SPR. This proves PSPACE-hardness. Membership in PSPACE follows from Savitch's Theorem [19] which states that PSPACE = NPSPACE; the problem is easily seen to be in NPSPACE. □

## 4 Chordal graphs

We will show in this section that for chordal graphs $G$, the SPR problem can be decided in polynomial time. In fact, we prove that if $G$ is chordal, then $\mathrm{SP}(G, s, t)$ is connected and has diameter at most $d - 1$, where $d$ is the distance from $s$ to $t$.

**Theorem 4 (*)** *Let $G$ be a chordal graph, and let $P$ and $Q$ be two shortest $st$-paths in $G$, of length $d$. Then a rerouting sequence from $P$ to $Q$ exists, of length at most $|V(P) \backslash V(Q)| \le d - 1$.*

*Proof sketch:* We prove the statement by induction over $c = |V(P) \backslash V(Q)|$. The case $c = 0$ is obvious, so now assume that $c \ge 1$. Let $P = u_0, u_1, \ldots, u_d$, and $Q = v_0, v_1, \ldots, v_d$. Let $i$ be the lowest index such that $u_i \ne v_i$. Let $j$ be the lowest index with $j > i$ such that $u_j = v_j$. If $j = i + 1$, then both $u_i$ and $v_i$ are adjacent to both $u_{i-1}$ and $u_{i+1}$, so to $P$ we can apply the rerouting step $u_i \to v_i$, to obtain a new shortest $st$-path $P'$ in $G$ that has one more vertex in common with $Q$. So by induction, the distance from $P'$ to $Q$ in $\mathrm{SP}(G, s, t)$ is at most $c - 1$. Then the distance from $P$ to $Q$ is at most $c$. This proves the statement in the case $j = i + 1$, so now assume that $j \geq i + 2$.

Note that then $C = u_{i-1}, u_i, \ldots, u_{j-1}, u_j, v_{j-1}, \ldots, v_i, v_{i-1}$ is a cycle in $G$, of length at least 6. Using $C$ and using that $G$ is chordal, it can be shown that $G$ contains an edge $e$ with $e = u_i v_{i+1}$ or $e = u_{i+1} v_i$. If $e = u_{i+1} v_i$, then both $u_i$ and $v_i$ are adjacent to both $u_{i-1}$ and $u_{i+1}$, so to $P$ we may apply the rerouting step $u_i \to v_i$, to obtain a new shortest $st$-path $P'$ that has at least one more vertex in common with $Q$. Then the proof can be concluded the same as before. The remaining case where $e = u_i v_{i+1}$ is symmetric; a rerouting step $v_i \to u_i$ can be applied to $Q$. □

The above proof gives a polynomial time algorithm for constructing the rerouting sequence. Obviously a rerouting sequence from $P$ to $Q$ requires at least $|V(P) \backslash V(Q)|$ rerouting steps, so we may conclude:

**Corollary 5** *Let $G$ be a chordal graph with shortest $st$-paths $P$ and $Q$. In polynomial time, a shortest rerouting sequence from $P$ to $Q$ can be constructed.*

## 5 Claw-free graphs

In this section we show that deciding SPR, and deciding whether $\mathrm{SP}(G, s, t)$ is connected can both be done in polynomial time in the case where $G$ is claw-free. A *claw* is a $K_{1,3}$ graph. A graph $G$ is *claw-free* if it contains no claws as induced subgraphs. In other words, $G$ is not claw-free if and only if it contains a subgraph $H$ that consists of one vertex $c$ of degree 3, and three leaves $l_1, l_2, l_3$, such that the leaves are pairwise nonadjacent in $G$. Such an *induced* subgraph will be called a *c-claw with leaves $l_1, l_2, l_3$* for short.

Let $u \in L_i$. We say that $u$ *has maximal in-neighborhood* if there is no $v \in L_i$ with $N^-(u) \subset N^-(v)$. (Note that we distinguish between subset $\subseteq$ and strict subset $\subset$.) In that case, $N^-(u)$ is a *maximal in-neighborhood in $L_{i-1}$*. These notions are defined analogously for out-neighborhoods. With a layer $L_i$, we associate the following hypergraph $\mathcal{H}_i$: $\mathcal{H}_i$ has vertex set $L_i$, and the edges correspond to the maximal in-neighborhoods in $L_i$. So for every $e \in E(\mathcal{H}_i)$, there exists a vertex $a \in L_{i+1}$ with $N^-(a) = e$.

The main result of this section is proved as follows. We first give some simple reduction rules. These are based on the fact that it is safe to delete a vertex $v$, if we know that it is not part of any shortest $st$-path that can be reached from the given shortest $st$-path $P$. We give two ways to identify such vertices. For reduced, claw-free SPR instances $G', P, Q$ that do not have such vertices, we actually show that $\mathrm{SP}(G', s, t)$ is connected. Proposition 6 follows from this observation: whenever a rerouting step $x \to y$ in layer $L_i$ is made, there is a vertex $z \in L_{i+1}$ with $x, y \in N^-(z)$, so $x$ and $y$ are in the same component of $\mathcal{H}_i$.

**Proposition 6** *Let $P$ be a shortest st-path in a graph $G$. For every shortest st-path $Q$ that is reachable from $P$ in $SP(G, s, t)$ and every $i$, the $L_i$-vertex is part of the same component of $\mathcal{H}_i$ as the $L_i$-vertex of $P$.*

**Proposition 7** *Let $P$ be a shortest st-path of length $d$ in a claw-free graph $G$, in which every vertex lies on a shortest st-path. For every shortest st-path $Q$ that is reachable from $P$ in $SP(G, s, t)$ and every $i \in \{2, \ldots, d-2\}$, the $L_i$-vertex of $Q$ is adjacent to the $L_i$-vertex of $P$.*

*Proof:* Consider a rerouting sequence $Q_0, \ldots, Q_k$ from $Q_0 = P$ to $Q_k = Q$, and let $x_j$ be the $L_i$-vertex of $Q_j$, for every $j \in \{0, \ldots, k\}$. Assume that the claim is not true, so then we may choose $\ell$ to be the lowest index such that $x_0 x_\ell \notin E(G)$.

If $x_0$ and $x_\ell$ have a common neighbor $z$ in either $L_{i-1}$ or $L_{i+1}$, then a $z$-claw with leaves $x_0, x_\ell$ and $y$ exists, for some vertex $y \in L_{i-2}$ or $y \in L_{i+2}$, respectively. ($z$ has such a neighbor $y$ since it lies on a shortest st-path.) So since $G$ is claw-free, we may conclude that $N^-(x_0) \cap N^-(x_\ell) = \emptyset$, and $N^+(x_0) \cap N^+(x_\ell) = \emptyset$. If $x_{\ell-1}$ has a neighbor $y \in L_{i-1} \setminus N^-(x_0)$ and a neighbor $z \in L_{i+1} \setminus N^+(x_0)$, then an $x_{\ell-1}$-claw with leaves $x_0, y, z$ exists. So w.l.o.g. we may assume that $N^-(x_{\ell-1}) \subseteq N^-(x_0)$. But then $N^-(x_{\ell-1}) \cap N^-(x_\ell) = \emptyset$, which contradicts that a rerouting step $x_{\ell-1} \to x_\ell$ is possible. □

**Definition 8** *Let $G$ be a claw-free graph with vertices $s$ and $t$ at distance $d$ from each other. Then $G$ is called st-reduced if*
- *all vertices lie on a shortest st-path,*
- *for every $i \in \{1, \ldots, d-1\}$, $\mathcal{H}_i$ is connected, and*
- *for every $i \in \{2, \ldots, d-2\}$, $L_i$ is a clique.*

Propositions 6 and 7 can be used to identify in polynomial time vertices that are not part of any shortest st-path that is reachable from a given path $P$. Deleting these, together with vertices not on shortest st-paths, gives an st-reduced instance.

**Lemma 9 (*)** *Let $G$ be a claw-free graph, with shortest st-path $P$. In polynomial time, we can construct an induced claw-free subgraph $G'$ such that (i) $G'$ is st-reduced, and (ii) a shortest st-path $Q$ of $G$ is reachable from $P$ in $SP(G, s, t)$ if and only if $V(Q) \subseteq V(G')$, and $Q$ is reachable from $P$ in $SP(G', s, t)$.*

The last property from Definition 8 shows that every pair of vertices in one layer is adjacent; this makes it much easier in our proofs to obtain a contradiction by exhibiting an induced claw. We use this to prove Lemmas 10 and 11.

**Lemma 10** *Let $P$ be a shortest st-path of length $d$ in a claw-free st-reduced graph $G$. In polynomial time, a rerouting sequence of length at most $d-1$ can be constructed, from $P$ to a shortest st-path $P'$ in which every vertex has maximal out-neighborhood. The same holds for the case of maximal in-neighborhoods.*

*Proof:* Let $P = u_0, u_1, \ldots, u_{d-1}, u_d$. Define $v_0 := u_0 (= s)$. For $i = 1, \ldots, d-1$, in increasing order, we change the $L_i$-vertex $u_i$ of $P$ as follows. If the out-neighborhood of $u_i$ is not maximal, then choose $v_i \in L_i$ with $N^+(u_i) \subset N^+(v_i)$, and $N^+(v_i)$ maximal. If possible, choose $v_i$ such that $v_i \in N^+(v_{i-1})$. Then, apply the rerouting step $u_i \to v_i$. If $u_i$ already has maximal out-neighborhood then simply define $v_i = u_i$.

It remains to show that $u_i \to v_i$ is in fact a rerouting step. By definition, $u_{i+1} \in N^+(v_i)$, so the $L_{i+1}$-vertex of the current path $v_0, \ldots, v_{i-1}, u_i, u_{i+1}, \ldots, u_d$ poses no problem. It might however be that $v_i$ is not adjacent to $v_{i-1}$. In that case, $i \geq 2$. Choose a vertex $x \in N^-(v_i)$. Since $v_i \in N^+(x) \setminus N^+(v_{i-1})$, but $N^+(v_{i-1})$ is maximal, there exists at least one $y \in N^+(v_{i-1}) \setminus N^+(x)$. By choice of $v_i$, there exists at least one $z \in N^+(v_i) \setminus N^+(y)$, otherwise $y$ has maximal out-neighborhood as well, and we would have chosen $v_i = y$ (since we gave preference to out-neighbors of $v_{i-1}$). This however gives a $v_i$-claw with leaves $x, y, z$, a contradiction. The in-neighborhood case is analog. □

Maximal in- and out-neighborhoods are required to apply the next lemma, which is concerned with rerouting a single layer $L_i$.

**Lemma 11 (*)** *Let $G$ be a claw-free, st-reduced graph, with distance $d$ between $s$ and $t$. Let $P = u_0, \ldots, u_d$ be a shortest st-path. Let $i \in \{1, \ldots, d-1\}$ such that $u_{i-1}$ has maximal out-neighborhood and $u_{i+1}$ has maximal in-neighborhood. Then for every $w \in N^+(u_{i-1})$, using at most $2|L_i|$ rerouting steps, $P$ can be modified to a shortest st-path $P' = v_0, \ldots, v_d$ with $v_i = w$, and $v_j = u_j$ for all $j \in \{0, \ldots, d\} \setminus \{i, i+1\}$.*

*Proof sketch:* We assume $1 \leq i \leq d-2$, since the case $i = d-1$ is trivial. Consider a shortest path $x_0, \ldots, x_k$ in $\mathcal{H}_i$ from $u_i$ to $w$, so $x_0 = u_i$ and $x_k = w$. (This exists since $G$ is $st$-reduced.) For $j \in \{1, \ldots, k\}$, let $a_j \in L_{i+1}$ be a vertex with maximal in-neighborhood such that $\{x_{j-1}, x_j\} \in N^-(a_j)$. (Such a vertex exists by the definition of the hypergraph $\mathcal{H}_i$.) Choose $a_1 = u_{i+1}$ if $u_{i+1}$ satisfies this condition. In addition, let $a_0 = u_{i+1}$. The rerouting sequence from $P$ to $P'$ uses the following rerouting steps: If $a_0 \neq a_1$, then first replace $a_0 = u_{i+1}$ by $a_1$. Next, replace $x_0 = u_i$ by $x_1$. Next, replace $a_1$ by $a_2$, and $x_1$ by $x_2$. Continue making rerouting steps alternatingly in layer $L_{i+1}$ and $L_i$ until $x_{k-1}$ can be replaced by $x_k = w$, to obtain the desired path $P'$. Note that by definition of the $x_j$ and $a_j$ vertices, at every point the $L_i$- and $L_{i+1}$-vertex are adjacent. It may however be that at some point, a vertex $x_j$ is not adjacent to $u_{i-1}$. In this case, the aforementioned rerouting sequence is preceded by replacing $u_{i-1}$ by a vertex $y \in N^-(x_j)$ with maximal out-neighborhood, and succeeded by replacing $y$ by $u_{i-1}$ again. Using the fact that $G$ is claw-free and $st$-reduced, it can be shown that this way, a shortest $st$-path is maintained throughout. □

Combining Lemmas 10 and 11 gives the main combinatorial result, for $st$-reduced claw-free graphs. The main algorithmic results, Theorem 14 and 13, follow easily. Their proofs are similar.

**Theorem 12** *Let $G$ be an st-reduced claw-free graph on $n$ vertices, with distance $d$ from $s$ to $t$. Between any two shortest $st$-paths $P$ and $Q$ in $G$, a rerouting sequence of length at most $2n + 2d - 6$ exists.*

*Proof:* First apply at most $d-1$ rerouting steps to $P$ to obtain a shortest $st$-path $P'$ in which every vertex has a maximal in-neighborhood (Lemma 10). Similarly, apply at most $d-1$ rerouting steps to $Q$ to obtain a shortest $st$-path $Q'$ in which every vertex has a maximal out-neighborhood (Lemma 10).

Now $P'$ can be modified to $Q'$ in $d-1$ stages $i$, with $i \in \{1, \ldots, d-1\}$. Denote $P_0 = P' = u_0, \ldots, u_d$, and $Q' = v_0, \ldots, v_d$. At the start of the $i$th stage, we have a shortest $st$-path $P_{i-1} = v_0, \ldots, v_{i-1}, a, u_{i+1}, \ldots, u_d$ for some $a \in L_i$ (note that for $i = 1$, $P_0$ is of this form). Using at most $2|L_i|$ rerouting steps, $P_{i-1}$ can be modified into a shortest $st$-path $P_i = v_0, \ldots, v_i, a', u_{i+2}, \ldots, u_d$ for some $a' \in L_{i+1}$. This follows from Lemma 11.

After $d-1$ stages, this procedure terminates with a path $v_0, \ldots, v_{d-1}, u_d$, which equals $Q'$. The total number of rerouting steps for these stages is at most $\sum_{i \in \{1, \ldots, d-1\}} 2|L_i| = 2(n-2)$. In total, this shows that $P$ and $Q$ can both be rerouted to a common shortest $st$-path $Q'$, in at most $2(n-2) + (d-1)$ and $d-1$ steps, respectively. Combining these rerouting sequences gives a rerouting sequence from $P$ to $Q$ of length at most $2n + 2d - 6$. □

**Theorem 13** *Let $G$ be a claw-free graph on $n$ vertices, and let $P$ and $Q$ be two shortest $st$-paths in $G$, of length $d$. In polynomial time it can be decided whether $Q$ is reachable from $P$ in $SP(G, s, t)$, and if so, a rerouting sequence of length at most $2n + 2d - 6$ exists.*

*Proof:* By Lemma 9, in polynomial time we can construct an $st$-reduced induced subgraph $G'$ of $G$ such that any shortest $st$-path $Q'$ is reachable from $P$ in $G$ if and only if it is reachable from $P$ in $G'$. So if $Q$ is not a shortest $st$-path of $G'$ (at least one of its vertices was deleted), we may conclude it is not reachable. Otherwise, Theorem 12 shows that $Q$ is reachable from $P$, with a rerouting sequence of length at most $2|V(G')| + 2d - 6 \leq 2n + 2d - 6$. □

**Theorem 14 (*)** *Let $G$ be a claw-free graph on $n$ vertices. In polynomial time it can be decided whether $SP(G, s, t)$ is connected.*

## 6 Isolated states

In this section we give a polynomial time algorithm for counting the number of isolated paths. Recall that an *isolated $st$-path* is a shortest $st$-path in $G$ that has no neighbors in $SP(G, s, t)$. For this, we need to consider isolated $sy$-paths for vertices $y \neq t$. For three vertices $s, x, y$ with $s \neq y$, we use $\text{iso}_{sy}(x)$ to denote the number of isolated $sy$-paths that contain the vertex $x$. We will use this notation for the case where $x$ is the second-to-last vertex on a shortest $sy$-path, so it is adjacent to $y$.

**Proposition 15** *Let $y$ and $z$ be vertices at distance $i$ and $i+1$ of $s$, respectively, with $i \geq 1$. Then $\text{iso}_{sz}(y) = \sum_x \text{iso}_{sy}(x)$, where the summation is over all vertices $x$ at distance $i-1$ from $s$ such that $N(x) \cap N(z) = \{y\}$.*

*Proof:* Let $x$ be such a vertex. There are $\text{iso}_{sy}(x)$ isolated $sy$-paths that end with $x$ and $y$. Since $y$ is the only common neighbor of $x$ and $z$, extending every one of these paths with the vertex $z$ gives a set of isolated $sz$-paths that contain $y$, which are all distinct. All of these paths have $x$ as $L_{i-1}$-vertex, so when choosing a different $L_{i-1}$-vertex in the role of $x$, a different set of $sz$-paths is obtained. This shows that there are at least $\sum_x \text{iso}_{sz}(y)$ distinct isolated $sz$-paths that contain $y$, where the summation is over all vertices $x$ at distance $i-1$ from $s$ such that $N(x) \cap N(z) = \{y\}$.

We conclude the proof by observing that every isolated $sz$-path that contains $y$ also contains some vertex $x$ at distance $i-1$ from $s$ with $N(x) \cap N(z) = \{y\}$, such that removing $z$ yields an isolated $sy$-path. Therefore, all isolated $sz$-paths that contain $y$ are counted this way, and thus we have equality. □

**Theorem 16** *Let $G$ be a graph with $s, t \in V(G)$. In polynomial time, the number of isolated $st$-paths can be computed.*

*Proof:* Let $d$ be the distance from $s$ to $t$. As usual, the layers $L_i$ for $i \in \{0, \ldots, d\}$ are defined with respect to $s$ and $t$. The algorithm works by computing the values $\text{iso}_{sz}(y)$ for various choices of $y$ and $z$, in increasing distance from $s$ to $z$. First, for every $z \in L_1$, initialize $\text{iso}_{sz}(s) = 1$ (which is trivially correct). Then, for $i = 2, \ldots, d$, in increasing order of $i$, do the following. For every $z \in L_i$ and every $y \in L_{i-1}$, compute $\text{iso}_{sz}(y)$ using Proposition 15. Note that the required values $\text{iso}_{sy}(x)$ have all been computed earlier. In the end, return $\sum_{y \in L_{d-1}} \text{iso}_{st}(y)$, which is the total number of isolated $st$-paths.

Now we analyze the complexity. Let $n = |V(G)|$. Note that the number of combinations of $y$ and $z$ for which $\text{iso}_{sz}(y)$ is computed is less than $n^2$. For every such combination, evaluating the expression from Proposition 15 can be done in polynomial time. □

The following example shows that this counting result is nontrivial: if we choose $G''$ to be just the main strand of the instance $G'$ constructed in Section 3 (based on a graph $G$ on $n$ vertices), then $G''$ contains $4^n$ isolated paths, which is exponential in the number of vertices of $G''$ (which is $14n + 2$).

## 7 Discussion

We showed that SPR is PSPACE-complete, which is somewhat surprising since the problem of finding shortest paths is easy. Nevertheless, our results otherwise confirm the typical behavior of reconfiguration problems: for instances where we can decide SPR in polynomial time (chordal and claw-free graphs), the diameter is polynomially bounded – in this case, even linearly bounded. In addition, for these graph classes it can be decided efficiently whether $SP(G, s, t)$ is connected.

The main question that is left open here is: *What is the complexity of deciding whether $SP(G, s, t)$ is connected, for general graphs $G$?*

We note that for the SPR instances $G'$, $P_\alpha$, $P_\beta$ constructed in Section 3, the proof of Lemma 2 shows that $SP(G', s, t)$ is always disconnected (unless the 4-Color Path instance $G$ has no edges).

We showed that for chordal graphs $G$, one can even find *shortest* rerouting sequences in polynomial time. Is this possible for claw-free graphs as well? To be precise, for two shortest $st$-paths $P$ and $Q$ in a claw-free graph $G$ and $k \in \mathbb{N}$, can it be decided in polynomial time whether a rerouting sequence from $P$ to $Q$ of length at most $k$ exists, or is this problem NP-complete? Recall that for general graphs, the NP-hardness of finding a shortest rerouting sequence was proved in [15]. By our linear diameter result, this (decision) problem lies in NP for claw-free graphs. Finally, it is interesting to search for other graph classes for which SPR can be solved in polynomial time. Graphs of bounded treewidth form a prime candidate.

**Acknowledgements** The author would like to thank Paul Medvedev and Martin Milanič for the introduction to this problem, and for comments on an earlier version of this paper. This research was supported by DFG grant BO 3391/1-1.

## A  Detailed proofs for Section 3

**Lemma 1** *If there is a recoloring sequence for $G$ from $\alpha$ to $\beta$, then there is a rerouting sequence from $P_\alpha$ to $P_\beta$ for $G'$.*

*Proof:* It suffices to show that for any two adjacent colorings $\gamma$ and $\delta$ in $\mathcal{C}_k(G)$, there is a rerouting sequence from $P_\gamma$ to $P_\delta$, where $P_\gamma$ and $P_\delta$ are the shortest $st$-paths in $G'$ that are constructed using $\gamma$ and $\delta$ as defined in Section 3. If this can be done for every consecutive pair in the recoloring sequence from $\alpha$ to $\beta$, then a rerouting sequence from $P_\alpha$ to $P_\beta$ exists. So, let $\gamma$ and $\delta$ be adjacent colorings in $\mathcal{C}_k(G)$. Let $v_i$ be the unique vertex in which they differ, and w.l.o.g. assume that $\gamma(v_i) = c_1$ and $\delta(v_i) = c_2$.

$P_\gamma$ can be transformed into $P_\delta$ as follows. See Figure 3, at the very end of the appendix, which continues on the example of Figure 2. Let $H_j^*$ denote the gadgets of the $v_i, \{c_1, c_2\}$-strand of $G'$.

First, for the layers $d = 1, \ldots, 5i - 3$, replace the vertex $v$ of $P_\gamma$ in layer $L_d$ by the unique vertex of $H_j^*$ (in the same layer) that has the same color as $v$. This is possible by making the changes in increasing layer order. Similarly, for the layers $d = 5n, 5n - 1, \ldots, 5i - 1$, replace the vertex $v$ in layer $L_d$ by the vertex of $H_j^*$ with the same color as $v$. This is possible by making the changes in decreasing layer order. This gives the path shown in Figure 3 (b). Note that these changes are possible if and only if $P_\gamma$ does not use vertices of color $c_1$ or $c_2$ from gadgets $H_j$ with $v_iv_j \in E(G)$. The latter property is ensured by the construction of $P_\gamma$ and $P_\delta$, since both $\gamma$ and $\delta$ are colorings of $G$, so all neighbors $v_j$ of $v_i$ have $\gamma(v_j) = \delta(v_j) \notin \{c_1, c_2\}$.

Now we can change the middle vertex of $H_i$ that is used in the path: replace the middle vertex of color $c_1$ with the one of color $c_2$, see Figure 3 (c). Next, we can move the entire path from the $v_i, \{c_1, c_2\}$-strand back to the main strand, similar to before (but in reverse order). See Figure 3 (d).

This yields a rerouting sequence from $P_\gamma$ to $P_\delta$. Since we can do this for every recoloring step in the recoloring sequence (there is a strand for every $v_i$ and every $\{c_1, c_2\}$), this concludes the proof. $\square$

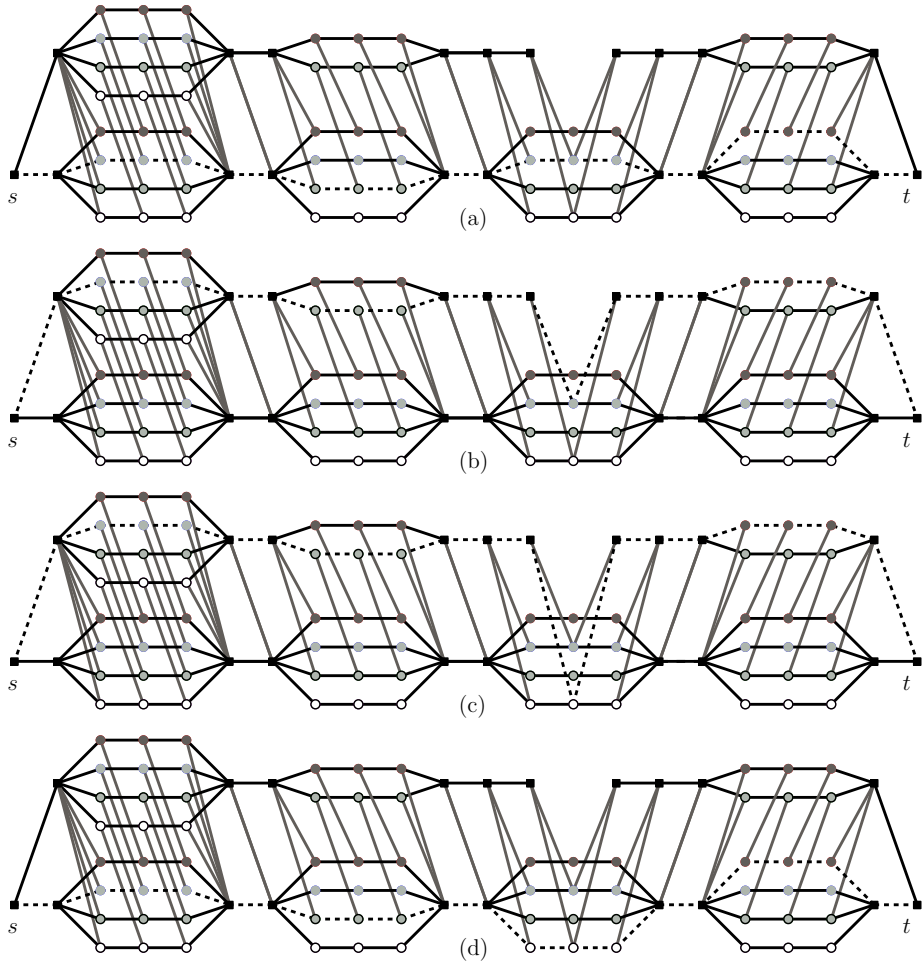

**Fig. 3.** Paths from a rerouting sequence from $P_\alpha$ to $P_\beta$.

**Lemma 2** *If there is a rerouting sequence from $P_\alpha$ to $P_\beta$ for $G'$, then there is a recoloring sequence in $G$ from $\alpha$ to $\beta$.*

*Proof:* First we define how *any* shortest $st$-path $P$ in $G'$ is mapped to a color assignment of $G$: $v_i$ receives the same color as the vertex of $P$ in layer $L_{5i-2}$; this is the layer that contains the middle vertices of $H_i$. Note that this defines a color assignment for $G$, but that this is not necessarily a (proper) coloring.

In a rerouting sequence from $P_\alpha$ to $P_\beta$, consider a step where the sequence moves from a path $P$ that corresponds to a color assignment $\gamma$, to a path $P'$ that corresponds to a different color assignment $\delta$. We will prove that if $\gamma$ is a coloring of $G$, then $\delta$ is a coloring of $G$ as well. (So then $\gamma$ and $\delta$ are adjacent vertices in $\mathcal{C}_k(G)$, since a rerouting step changes at most one color.) Say $\gamma$ and $\delta$ differ in $v_i$, where $\gamma(v_i) = c_1$ and $\delta(v_i) = c_2$.

First we observe that the path $P$ contains the $l$ and $r$ vertices of the $v_i, \{c_1, c_2\}$-strand: $l$ is the only vertex in layer $L_{5i-3}$ that is adjacent to both a vertex of color $c_1$ in $L_{5i-2}$ and a vertex of color $c_2$ in $L_{5i-2}$. Similarly, $r$ is the only such vertex in layer $L_{5i-1}$.

Therefore, all vertices of $P$ except the one in layer $L_{5i-2}$ are part of the $v_i, \{c_1, c_2\}$-strand. Indeed, the only neighbor of $l$ in layer $L_{5i-4}$ is part of this strand (this is $s_i^*$), and the only neighbor of $s_i^*$ in layer $L_{5i-5}$ is part of this strand (this is $t_{i-1}^*$), all neighbors of $t_{i-1}^*$ in layer $L_{5i-6}$ are part of this strand, etc. Similarly, starting from $r$ we can argue that the vertices of $P$ in layers $5i$, $5i+1$, etc. are part of this strand.

Since we now have that all internal vertices of $P$ lie in the $v_i, \{c_1, c_2\}$-strand, we conclude that for all for neighbors $v_j$ of $v_i$, $\gamma(v_j) \in \{1,2,3,4\}\backslash\{c_1, c_2\}$: This follows from the construction of the $v_i, \{c_1, c_2\}$-strand (recall that for neighbors $v_j$ of $v_i$, $H_j^*$ contains no vertices of color $c_1$ or $c_2$).

So if the coloring $\gamma$ is modified by changing the color of $v_i$ from $\gamma(v_i) = c_1$ to $\delta(v_i) = c_2$, then again a coloring of $G$ is obtained, which is $\delta$.

We conclude that all paths in the rerouting sequence correspond to colorings, since we started with one that corresponded to a coloring, namely $P_\alpha$. □

## B  Proof details for Theorem 4

Proposition 17 provides the missing step in the proof of Theorem 4, as given in the main text. For completeness, subsequently we give the full proof of Theorem 4 again.

**Proposition 17** *Let $C = v_0, v_1, \ldots, v_k$ be a cycle in a chordal graph $G$, of length $k \geq 4$. Let $v_i v_{i+1}$ and $v_j v_{j+1}$ be edges of $C$ that share no end vertices, with $i < j$. Then $v_i v_j$ or $v_{i+1} v_{j+1}$ is a chord of $C$.*

*Proof:* Consider a shortest cycle $C'$ that contains both of the edges $e = v_i v_{i+1}$ and $f = v_j v_{j+1}$, such that $v_{i+1}$ follows $v_i$ and $v_{j+1}$ follows $v_j$ in the vertex sequence. (Considering $C$, we know that such a cycle $C'$ exists.) Since $e$ and $f$ share no end vertices, $C'$ has length at least 4. Because $G$ is chordal, $C'$ has a chord $g$. If this chord $g$ is not equal to $v_i v_j$ or $v_{i+1} v_{j+1}$, then it can be seen that $g$ can be combined with a part of $C'$ to find a shorter cycle that contains both $e$ and $f$, where $v_{i+1}$ follows $v_i$ and $v_{j+1}$ follows $v_j$. But that contradicts the choice of $C'$, so $g$ must be one of the aforementioned chords. □

**Theorem 4** *Let $G$ be a chordal graph, and let $P$ and $Q$ be two shortest st-paths in $G$, of length d. Then a rerouting sequence from $P$ to $Q$ exists, of length at most $|V(P)\backslash V(Q)| \leq d - 1$.*

*Proof:* We prove the statement by induction over $c = |V(P)\backslash V(Q)|$. If $c = 0$ then $P = Q$, so the statement is trivial. So now assume that $c \geq 1$; $P$ and $Q$ differ in at least one vertex. Let $P = u_0, u_1, \ldots, u_d$, and $Q = v_0, v_1, \ldots, v_d$. Let $i$ be the lowest index such that $u_i \neq v_i$ (such an $i$ exists since $c \geq 1$). Let $j$ be the lowest index with $j > i$ such that $u_j = v_j$ (since $u_d = t = v_d$, such a $j$ exists). If $j = i + 1$, then both $u_i$ and $v_i$ are adjacent to both $u_{i-1}$ and $u_{i+1}$, so to $P$ we can apply the rerouting step $u_i \to v_i$, to obtain a new shortest $st$-path $P'$ in $G$ that has one more vertex in common with $Q$. So by induction, the distance from $P'$ to $Q$ in $\mathrm{SP}(G,s,t)$ is at most $c - 1$. Then the distance from $P$ to $Q$ is at most $c$. This proves the statement in the case $j = i + 1$, so now assume that $j \geq i + 2$.

Note that then $C = u_{i-1}, u_i, \ldots, u_{j-1}, u_j, v_{j-1}, \ldots, v_i, v_{i-1}$ is a cycle in $G$, of length at least 6. By applying Proposition 17 to $C$ and the edges $u_i u_{i+1}$ and $v_{i+1} v_i$, we may conclude that $G$ contains an edge $e$ with $e = u_i v_{i+1}$ or $e = u_{i+1} v_i$. If $e = u_{i+1} v_i$, then both $u_i$ and $v_i$ are adjacent to both $u_{i-1}$ and $u_{i+1}$, so to $P$ we may apply the rerouting step $u_i \to v_i$, to obtain a new shortest $st$-path $P'$ that has at least one more vertex in common with $Q$. Then the proof can be concluded the same as before. In the remaining case, $e = u_i v_{i+1}$. Then both $u_i$ and $v_i$ are adjacent to both $v_{i-1}$ and $v_{i+1}$. So we can obtain a shortest $st$-path $Q'$ from $Q$ by changing the vertex $v_i$ to $u_i$. Then by induction, the distance from $P$ to $Q'$ is at most $c - 1$, so the distance from $P$ to $Q$ is at most $c$.

This covers all cases and concludes the induction step, so we may conclude that the distance from $P$ to $Q$ is at most $c$. □

## C  Proofs omitted from Section 5

**Lemma 9** *Let $G$ be a claw-free graph, with shortest st-path $P$. In polynomial time, we can construct an induced claw-free subgraph $G'$ such that*

(i) *$G'$ is st-reduced, and*
(ii) *a shortest st-path $Q$ of $G$ is reachable from $P$ in $\mathrm{SP}(G, s, t)$ if and only if $V(Q) \subseteq V(G')$, and $Q$ is reachable from $P$ in $\mathrm{SP}(G', s, t)$.*

*Proof:* Let $d$ denote the length of $P$. If we know that a given vertex $v$ is not part of any shortest $st$-path that can be reached from $P$ in $\mathrm{SP}(G,s,t)$, then it is easily seen that deleting $v$ is *safe*, i.e. $G'$ satisfies Property (ii) from the lemma statement.

To obtain $G'$ from $G$ we delete the following vertices. First, we delete every vertex that does not lie on a shortest $st$-path, which clearly is safe. Secondly, for every $i \in \{1, \ldots, d-1\}$, we delete the vertices that do not lie in the same component of $\mathcal{H}_i$ as the $L_i$-vertex of $P$. By Proposition 6, this is safe. Finally, for every $i \in \{2, \ldots, d-2\}$, we delete every vertex in $L_i$ that is not adjacent to the $L_i$-vertex of $P$. By Proposition 7, this is safe. Call the resulting graph $G'$. Clearly, $G'$ is an $st$-reduced graph. Since we only deleted vertices, $G'$ is an induced subgraph of $G$, and therefore again claw-free. $\square$

For proving Lemma 11 in detail, we require the following two properties of $st$-reduced graphs.

**Proposition 18** *Let $G$ be a claw-free, $st$-reduced graph, with distance $d$ from $s$ to $t$. For $i \in \{1, \ldots, d-1\}$, let $x_0, \ldots, x_\ell$ be a shortest path in $\mathcal{H}_i$. Then for all $j \in \{1, \ldots, \ell-1\}$ and $k \in \{0, \ldots, \ell\}$, it holds that $N^-(x_j) \subseteq N^-(x_k)$.*

*Proof:* Suppose to the contrary that there exists a vertex $y \in N^-(x_j) \setminus N^-(x_k)$, for some $j \in \{1, \ldots, \ell-1\}$ and $k \in \{0, \ldots, \ell\}$. W.l.o.g. we may assume that $k > j$. Let $a_0, \ldots, a_{k-1}$ be vertices in $L_{i+1}$ such that for all $p \in \{0, \ldots, k-1\}$, $\{x_p, x_{p+1}\} \subseteq N^-(a_p)$. By definition of $\mathcal{H}_i$, such vertices exist.

We now claim that there exists an $x_j$-claw, with leaves $y, a_{j-1}, x_k$. These three vertices are all adjacent to $x_j$ ($x_k$ is adjacent since $G$ is $st$-reduced so $L_i$ is a clique). Clearly, $y \in L_{i-1}$ and $a_{j-1} \in L_{i+1}$ are not adjacent. By choice of $y$, it is not adjacent to $x_k$. If $x_k \in N^-(a_{j-1})$, then a shorter path from $x_0$ to $x_k$ in $\mathcal{H}_i$ would exist, namely $x_0, \ldots, x_{j-1}, x_k, \ldots, x_\ell$, a contradiction. Hence this is indeed an induced claw, and thus from this contradiction we may conclude that for all $j, k$ as stated, $N^-(x_j) \subseteq N^-(x_k)$. $\square$

**Proposition 19** *Let $G$ be a claw-free and $st$-reduced graph. If $u, v \in L_i$ have distinct maximal in-neighborhoods, then $N^+(u) = N^+(v)$. If $u, v \in L_i$ have distinct maximal out-neighborhoods, then $N^-(u) = N^-(v)$.*

*Proof:* Suppose $u, v \in L_i$ have distinct maximal in-neighborhoods, and that there exists $x \in N^+(u) \setminus N^+(v)$. Then we may choose $y \in N^-(u) \setminus N^-(v)$. This gives a $u$-claw with leaves $x, y, v$. Note that $uv \in E(G)$ since $G$ is $st$-reduced. The other cases are analog. $\square$

Now we can prove Lemma 11 in detail.

**Lemma 11** *Let $G$ be a claw-free, $st$-reduced graph, with distance $d$ between $s$ and $t$. Let $P = u_0, \ldots, u_d$ be a shortest $st$-path. Let $i \in \{1, \ldots, d-1\}$ such that $u_{i-1}$ has maximal out-neighborhood and $u_{i+1}$ has maximal in-neighborhood. Then for every $w \in N^+(u_{i-1})$, using at most $2|L_i|$ rerouting steps, $P$ can be modified to a shortest $st$-path $P' = v_0, \ldots, v_d$ with $v_i = w$, and $v_j = u_j$ for all $j \in \{0, \ldots, d\} \setminus \{i, i+1\}$.*

*Proof:* If $i = d-1$, the proof is trivial: $u_{i+1} = t$, which is adjacent to every vertex in $L_{d-1}$. Therefore, we may simply apply the single rerouting step $u_i \to v$, to obtain the desired shortest $st$-path $P'$. So now assume that $1 \leq i \leq d-2$.

Consider a shortest path $x_0, \ldots, x_k$ in $\mathcal{H}_i$ from $u_i$ to $w$, so $x_0 = u_i$ and $x_k = w$. For $j \in \{1, \ldots, k\}$, let $a_j \in L_{i+1}$ be a vertex with maximal in-neighborhood such that $\{x_{j-1}, x_j\} \in N^-(a_j)$. (Such a vertex exists by the definition of the hypergraph $\mathcal{H}_i$.) Choose $a_1 = u_{i+1}$ if $u_{i+1}$ satisfies this condition. In addition, let $a_0 = u_{i+1}$.

The plan is to use this path to reroute $P$ to $P'$. In particular, the rerouting sequence that we construct below uses the rerouting steps $x_0 \to x_1 \to \ldots \to x_k$ in layer $L_i$. But it may be necessary to make changes in layers $L_{i-1}$ and $L_{i+1}$ as well.

Firstly, a rerouting step in $L_{i-1}$ is required if there exists a $j$ with $u_{i-1}x_j \notin E(G)$. In this case, let $y \in L_{i-1}$ be an in-neighbor of $x_j$ with maximal out-neighborhood. (Note that $x_j$ has at least one in-neighbor $y$ that has maximal out-neighborhood.) We claim that $u_{i-2}y \in E(G)$: this follows since $u_{i-1}$ has maximal out-neighborhood as well, so $N^-(u_{i-1}) = N^-(y)$ (Proposition 19). Secondly, Proposition 18 shows that for every $\ell$, $yx_\ell \in E(G)$. On the other hand, if $u_{i-1}$ is actually adjacent to every $x_j$, in particular if $i = 1$, then for the remainder of the proof we simply choose $y = u_{i-1}$. The rerouting sequence from $P$ to $P'$ is given by the following series of modifications (below we prove that these are all actual rerouting steps):

(1)     in $L_{i-1}$: $u_{i-1} \to y$    (Skip if $y = u_{i-1}$.)
(2)     in $L_{i+1}$: $a_0 \to a_1$    (Recall that $a_0 = u_{i+1}$. Skip if $a_0 = a_1$.)
(3)     in $L_i$:    $x_0 \to x_1$    (Recall that $x_0 = u_i$.)
(4)     in $L_{i+1}$: $a_1 \to a_2$
(5)     in $L_i$:    $x_1 \to x_2$

$$\vdots$$

$(2k)$    in $L_{i+1}$: $a_{k-1} \to a_k$
$(2k+1)$ in $L_i$:    $x_{k-1} \to x_k$ (Recall that $x_k = w$.)
$(2k+2)$ in $L_{i-1}$: $y \to u_{i-1}$    (Skip if $y = u_{i-1}$.)

Let $Q_0, \ldots, Q_m$ be the vertex sequences that result from these changes, starting with with $Q_0 = P$. We first verify that for every $\ell \in \{0, \ldots, m\}$, $Q_\ell$ is a shortest $st$-path. In other words, we show that for every $\ell$, the $\ell$th change $p \to q$ above is a rerouting step; we verify that the vertices of $Q_{\ell-1}$ in the previous and next layer are both also adjacent to $q$.

As observed above, if $i \geq 2$, then $y$ is adjacent to $u_{i-2}$, so in every $Q_\ell$ the $L_{i-2}$-vertex and $L_{i-1}$-vertex are adjacent. Furthermore, $y$ is adjacent to every $x_j$, so it is adjacent to the $L_i$-vertex of every $Q_\ell$. This shows that for every $Q_\ell$, the $L_{i-1}$-vertex and the $L_i$-vertex are adjacent.

Now we show that $L_i$-vertex and $L_{i+1}$-vertex are adjacent in every $Q_\ell$. If a rerouting step $x_j \to x_{j+1}$ is made in layer $L_i$, then at that point, the vertex in layer $L_{i+1}$ is $a_{j+1}$, which by definition is adjacent to both $x_j$ and $x_{j+1}$. Similarly, if a rerouting step $a_j \to a_{j+1}$ is made in layer $L_{i+1}$, then at that point the $L_i$-vertex is $x_j$, which is adjacent to both.

Finally, we show that the $L_{i+1}$-vertex and $L_{i+2}$-vertex are adjacent in every $Q_\ell$. We first argue that whenever a rerouting step $a_j \to a_{j+1}$ is applied, $a_j$ and $a_{j+1}$ have distinct maximal in-neighborhoods. For $j \geq 1$, this follows from the fact that $x_0, \ldots, x_k$ is a *shortest* path in $\mathcal{H}_i$, so there is no in-neighborhood that contains both $x_{j-1}$ and $x_{j+1}$. For $a_0$ and $a_1$ it follows from the choice of $a_1$: recall that $a_0 = u_{i+1}$, which we assumed to have a maximal in-neighborhood. So if $x_1 \notin N^-(a_0)$, then $a_0$ and $a_1$ again have distinct maximal in-neighborhoods. If $x_1 \in N^-(a_0)$, then we have chosen $a_1 = a_0$, and in fact no rerouting step is made. Hence we may now conclude that Proposition 19 can be applied for every rerouting step $a_j \to a_{j+1}$, which shows that $N^+(a_j) = N^+(a_{j+1})$ for every $j$. Therefore, $u_{i+2}$ is an out-neighbor of every $a_j$.

This concludes the proof that every $Q_\ell$ is a shortest $st$-path, so $Q_0, \ldots, Q_m$ is a rerouting sequence, which results in the path $Q_m = u_0, \ldots, u_{i-1}, w, a_k, u_{i+2}, \ldots, u_d$, which is of the form we required for $P'$. Finally, note that this rerouting sequence used at most $2k+2$ rerouting steps. Since $k \leq |L_i|-1$, this proves the statement. □

It remains to prove Theorem 14. The proof is similar to the proof of Theorem 13.

**Theorem 14** *Let $G$ be a claw-free graph on $n$ vertices. In polynomial time it can be decided whether $SP(G, s, t)$ is connected.*

*Proof:* In polynomial time we can first delete all vertices of $G$ that do not lie on a shortest $st$-path, to obtain $G'$. Clearly, $\mathrm{SP}(G, s, t) = \mathrm{SP}(G', s, t)$, and $G'$ is again claw-free. Choose an arbitrary shortest $st$-path $P$. Using $G'$ and $P$, Lemma 9 can be applied to obtain an $st$-reduced subgraph $G''$ of $G'$ in polynomial time. If $G'' = G'$ then Theorem 12 shows that $\mathrm{SP}(G'', s, t) = \mathrm{SP}(G', s, t)$ is connected. Otherwise, there exists at least one vertex $v \in V(G') \backslash V(G'')$, and we may conclude that $\mathrm{SP}(G', s, t)$ is not connected: $G'$ has a shortest $st$-path $Q'$ with $v \in V(Q')$, which is not part of $G''$, but all shortest $st$-paths that are reachable from $P$ are part of $G''$ (Lemma 9). □